\documentclass[twocolumn]{aa}
\usepackage{graphicx}
\usepackage{txfonts}
\def\lsim{\lower.5ex\hbox{$\; \buildrel < \over \sim \;$}}
\def\gsim{\lower.5ex\hbox{$\; \buildrel > \over \sim \;$}}
\begin{document}
\title{Is $^6$Li in metal-poor halo stars \\ produced in situ by 
solar-like flares?}

\author{V. Tatischeff \and J.-P. Thibaud}

\offprints{V. Tatischeff}

\institute{Centre de Spectrom\'etrie Nucl\'eaire et de Spectrom\'etrie de 
Masse, CNRS/IN2P3 and Univ Paris-Sud, F-91405 Orsay, France\\
              \email{tatische@csnsm.in2p3.fr}
%             \thanks{}
             }

\date{Received ...; accepted ...}

   \abstract{

The high $^6$Li abundances recently measured in metal-poor halo stars are far
above the value predicted by Big Bang nucleosynthesis. They cannot be explained
by galactic cosmic-ray interactions in the interstellar medium either. Various 
pre-galactic sources of $^6$Li have been proposed in the literature. We study 
the possibility that the observed $^6$Li was produced by repeated solar-like 
flares on the main sequence of these low-metallicity stars. The time-dependent
flaring activity of these objects is estimated from the observed evolution of 
rotation-induced activity in Pop~I dwarf stars. As in solar flares, $^6$Li 
could be mainly created in interactions of flare-accelerated $^3$He with 
stellar atmospheric $^4$He, via the reaction $^4$He($^3$He,$p$)$^6$Li. Stellar 
dilution and destruction of flare-produced $^6$Li are evaluated from the 
evolutionary models of metal-poor stars developed by Richard and co-workers.
Stellar depletion should be less important for $^6$Li atoms synthesized in 
flares than for those of protostellar origin. Theoretical frequency 
distributions of $^6$Li/$^7$Li ratios are calculated using a Monte-Carlo method 
and compared with the observations. Excellent agreement is found with the 
measured $^6$Li/$^7$Li distribution, when taking into account the contribution 
of protostellar $^6$Li originating from galactic cosmic-ray nucleosynthesis. We 
propose as an observational test of the model to seek for a positive 
correlation between $^6$Li/$^7$Li and stellar rotation velocity. We also show 
that the amounts of $^7$Li, Be and B produced in flares of metal-poor halo 
stars are negligible as compared with the measured abundances of these
species. $^6$Li in low-metallicity stars may be a unique evidence of 
the nuclear processes occuring in stellar flares.

   \keywords{stars: abundances -- stars: flare -- nuclear reactions, 
   nucleosynthesis, abundances}
   }

\titlerunning{Flare production of $^6$Li in metal-poor halo stars}

   \maketitle
%
%________________________________________________________________

\section{Introduction}

The origins of the light elements Li, Be, and B (hereafter LiBeB) differ from
those of heavier nuclides. Whereas most elements are produced by stellar
nucleosynthesis, LiBeB are mainly destroyed in stellar interiors by 
thermonuclear reactions with protons. Thus, $^6$Li is rapidly consumed at 
stellar temperatures higher than 2$\times$10$^6$~K. The major source of $^6$Li 
has been thought for decades to be the interaction of galactic cosmic rays
(GCRs) with the interstellar medium (for a review see Vangioni-Flam et al. 
2000). Unlike $^7$Li, $^6$Li is predicted to be formed at a very low level in 
Big Bang nucleosynthesis, $^6$Li/H$\simeq$10$^{-14}$ (Thomas et al. 1993; 
Vangioni-Flam et al. 1999).

However, recent major observational advances have allowed new measurements of 
the $^6$Li/$^7$Li isotopic ratio in stars, which challenge our understanding of 
the origin of $^6$Li. Asplund et al. (2006) have recently reported the 
observation of $^6$Li in several halo stars of low metallicity, [Fe/H]$<$-1 
(where [Fe/H]=$\log[{\rm (Fe/H)/(Fe/H)}_\odot]$ and (Fe/H)/(Fe/H)$_\odot$ is 
the Fe abundance relative to its solar value). These authors have detected 
$^6$Li at $\geq$2$\sigma$ confidence level in nine of 24 studied stars. The 
$^6$Li/H values measured in these metal-poor halo stars (MPHSs) are 
$>$10$^{-12}$, i.e. more than 100 times higher than the predicted abundance 
from standard Big Bang nucleosynthesis. Reported $^6$Li abundances at 
[Fe/H]$\lsim$-2.3 are also larger than expected if GCR nucleosynthesis was the 
major source of $^6$Li (Rollinde et al. 2005; Prantzos et al. 2006). This 
result was already pointed out by Ramaty et al.(2000a) after the first 
detection of $^6$Li in two MPHSs of metallicity [Fe/H]$\approx$-2.3 (Smith et 
al. 1993; Hobbs \& Thorburn 1997; Smith et al. 1998; Cayrel et al. 1999). 

Two alternative types of pre-galactic $^6$Li sources have been proposed: (1) 
production in the early universe induced by the decay of supersymmetric dark 
matter particles during Big Bang nucleosynthesis (Jedamzik 2000; Jedamzik et 
al. 2005; Kawasaki et al. 2005; Ellis et al. 2005; Kusakabe et al. 2006) and 
(2) production by the interaction of cosmological cosmic rays that could be 
accelerated in shocks induced by large-scale structure formation (Suzuki \& 
Inoue 2002) or by an early population of massive stars (Pop~III stars; 
Rollinde et al. 2005, 2006). 

Because a substantial stellar depletion of $^6$Li seems to be unavoidable, any 
model of pre-galactic production should account for a higher $^6$Li abundance 
in the early Galaxy than those measured in evolved MPHSs. The 
value of the so-called "Spite plateau" (Spite \& Spite 1982), namely the 
nearly constant abundance of Li (mostly $^7$Li) measured in most MPHSs (see 
Bonifacio et al. 2006), is a factor of two to four lower than the primordial 
$^7$Li abundance calculated in the framework of standard Big Bang 
nucleosynthesis ($^7$Li/H=4.27$_{-0.83}^{+1.02} \times 10^{-10}$, Cyburt 2004; 
$^7$Li/H=4.15$_{-0.45}^{+0.49} \times 10^{-10}$, Coc et al. 2004). According to 
recent stellar-evolution models that include diffusion and turbulent mixing
(Richard et al. 2005; Korn et al. 2006), this discrepancy is probably due to 
the stellar depletion of $^7$Li. In such a case, an even larger depletion of 
pre-galactic $^6$Li is expected. Given that depletion during the pre-main 
sequence of the MPHSs is predicted to be negligeable for $^7$Li, but 
$\geq$0.3~dex for $^6$Li (Richard et al. 2002, 2005; Piau 2005), the stellar 
destruction of protostellar $^6$Li is probably higher than 0.6~dex. 

In this paper, we assess the possibility that the excess $^6$Li discovered in
MPHSs is not of pre-galactic origin, but was produced {\it in situ} by repeated 
solar-like flares on the main sequence of these stars. Predictions for 
significant synthesis of $^6$Li in stellar flares have been made long ago 
(Canal et al. 1975; Walker et al. 1985; see also Ryter et al. 1970). $^6$Li 
enhancement has been found during a long-duration flare of a chromospherically 
active binary (Montes \& Ramsey 1998) and in the atmosphere of a single 
chromospherically active K dwarf (Christian et al. 2005). In both cases, a 
flare production was found to be consistent with the activity level of the 
object. A high Li abundance associated with a large flare has also been 
observed in a very active late-type dwarf star (Mathioudakis et al. 1995). In 
this case, however, the Li isotopic ratio was not measured. 

Evidence for significant synthesis of $^6$Li in large solar flares is 
provided by optical observations of sunspots (Ritzenhoff et al. 1997) and
measurements of the solar wind lithium isotopic ratio in lunar soil (Chaussidon
\& Robert 1999). Calculations of $^6$Li production by nuclear interaction of
solar-flare-accelerated particles were performed by Ramaty et al. (2000b). They
have shown that a major $^6$Li production channel is due to accelerated $^3$He
interactions with solar atmospheric $^4$He, via the reaction 
$^4$He($^3$He,$p$)$^6$Li. 

Deliyannis \& Malaney (1995) have already studied the {\it in situ} production 
of $^6$Li from flares of MPHSs, following the first report of a $^6$Li 
detection in the halo star HD~84937 (Smith et al. 1993). They found that large 
flares could account for the $^6$Li abundance measured in that star. However,
Lemoine et al. (1997) have questioned this result on the basis of the 
$^6$Li production efficiency: assuming the flaring activity of the contemporary 
Sun for one billion years, they found the amount of flare-produced $^6$Li to be
negligible as compared with the abundance measured in HD~84937.

There are two main reasons that flare production of $^6$Li in MPHSs can be more 
important than previously thought. First, the $^6$Li production efficiency was 
underestimated in previous studies, because the reaction 
$^4$He($^3$He,$p$)$^6$Li was not taken into account. Second, it is very likely 
that most halo stars were much more active in their youth than the contemporary 
Sun. Indeed stellar rotation being the decisive factor of solar-like activity 
of dwarf stars, magnetic braking during the main sequence leads to a well-known 
decay with stellar age of chromospheric and coronal activities (e.g. Gershberg 
2005). 

We present in this paper a time-dependent model for flare production of LiBeB 
in MPHSs. The flaring activity of these Population~II stars is estimated from 
the observed evolution of rotation-induced activity in their Pop~I 
counterparts. Dilution and depletion of flare-produced $^6$Li in the surface 
convection zone of MPHSs are evaluated from the evolutionary stellar models of 
Richard et al. (2002). In Sect.~2 we first calculate the efficiency for 
synthesis of $^6$Li and the other LiBeB isotopes in stellar flares. The model 
for the time evolution of $^6$Li abundance at the surface of MPHSs is 
described in Sect.~3.1. In Sect.~3.2 we use a Monte-Carlo simulation to 
calculate theoretical frequency distributions of $^6$Li/$^7$Li values in MPHSs, 
which are then compared to the observations. In Sect.~4.1 we comment on the 
dispersion of the $^6$Li/$^7$Li data in the light of our Monte-Carlo results. 
We discuss in Sect.~4.2 $^6$Li production from flares of young stellar 
objects. We show in Sect.~4.3 that $^6$Li is the only LiBeB isotope which can 
be produced in non-negligible amount in flares of MPHSs relative to other 
production modes. Observational tests of the flare model are discussed in 
Sect.~4.4. A summary is given in Sect.~5. 

\section{Yields of LiBeB production in stellar flares}
 
We calculated the efficiency for LiBeB production in flares of MPHSs assuming 
the same thick target interaction model that is usually employed to describe 
nuclear processes in solar flares (see Tatischeff et al. 2006). Fast particles 
with given energy spectra and composition are assumed to be accelerated in the 
stellar corona and to produce nuclear reactions as they slow down in the lower 
stellar atmosphere.  

For the source energy spectrum of the fast ions we used an unbroken power 
law of spectral index $s$, extending from the threshold energies of the various 
nuclear reactions up to $E_{max}$=1~GeV nucleon$^{-1}$. For solar flares, 
Ramaty et al. (1996) found from analyses of gamma-ray line ratios a range of 
spectral indexes of about $s$=4$\pm$1. We assumed that $s$$\simeq$4 is also
close to the mean of spectral index distribution in stellar flares. 

The ambient and accelerated-ion abundances are also based on those employed 
in solar flare studies (Tatischeff et al. 2006), but we rescaled the 
abundances of both ambient and fast C and heavier elements to the metallicity 
of the MPHSs. We performed calculations with an accelerated $\alpha$/$p$ ratio 
of 0.1, which is typical of the abundance ratio measured in impulsive solar 
energetic events (e.g. Reames 1999). Gamma-ray spectroscopic analyses have 
provided evidence that the accelerated $\alpha$/$p$ ratio could be as high as 
0.5 in some solar flares (Share \& Murphy 1997; Mandzhavidze et al. 1999). It 
is noteworthy that such a large $\alpha$/$p$ ratio would significantly enhance 
the efficiency for production of $^6$Li, because this isotope is mainly 
synthesized in He+He interactions (see below). However, we adopted here the
canonical $\alpha$/$p$=0.1.

We took the accelerated $^3$He/$\alpha$ ratio to be 0.5, which is also typical 
of the abundances found in impulsive solar energetic events (Reames 1999). Such
an enrichment of fast $^3$He is caused by resonant wave-particle processes that 
are characteristic of the stochastic acceleration mechanism at work in
impulsive solar flares (e.g. Temerin \& Roth 1992). We assume that this
acceleration process enhances the accelerated $^3$He in stellar flares as well. 

The cross sections for the nuclear reactions are mostly from Ramaty et al. 
(1997), but we took into account the more recent measurements of Mercer 
et al. (2001) for production of $^6$Li and $^7$Li in the $\alpha + \alpha$ 
reaction. The cross section for the reaction $^4$He($^3$He,$p$)$^6$Li was
evaluated by Ramaty et al. (2000b). We also included in our calculations the 
reaction $^{12}$C($^3$He,$x$)$^7$Be ($^7$Be decays to $^7$Li with a half-life 
of 53~days), whose cross section was evaluated by Tatischeff et al. (2006).

   \begin{figure}
   \centering
   \includegraphics[width=8.5cm]{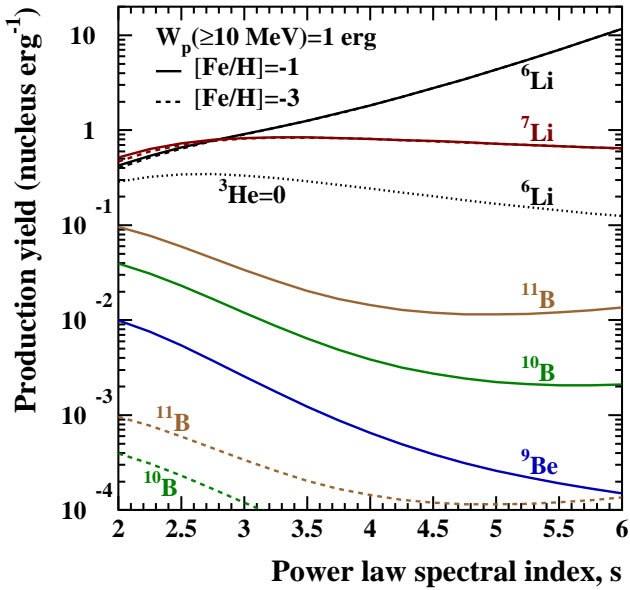}
      \caption{LiBeB productions by stellar-flare accelerated particles with 
      power law in kinetic energy per nucleon spectra with spectral index $s$. 
      The calculations are normalized to a total kinetic energy of 1 erg in 
      protons of energy $E$$\ge10~{\rm MeV}$. {\it Solid curves}: [Fe/H]=-1. 
      {\it Dashed curves}: [Fe/H]=-3. Li production is almost independent 
      of metallicity, because it is mostly due to accelerated $^3$He
      and $\alpha$-particle interactions with ambient He. {\it Dotted curve}: 
      $^6$Li production for $^3$He=0 and [Fe/H]=-1. 
              }
         \label{Fig1}
   \end{figure}

Figure~1 shows calculated thick-target yields for LiBeB productions as a
function of spectral index $s$. The calculations are normalized to a total
kinetic energy of 1 erg contained in flare-accelerated protons of energy 
greater than 10~MeV. We see that for $s$$>$2.8, the largest production is that 
of $^6$Li. The importance of the reaction $^4$He($^3$He,$p$)$^6$Li for the 
synthesis of this isotope can be seen by comparing the upper curve with the
dotted one, for which we set the accelerated $^3$He abundance to zero. For 
$s$=4, the $^3$He+$^4$He reaction accounts for 87\% of the total $^6$Li
production. 

The metallicity dependence of the production yields can be seen by comparing 
the curves obtained for [Fe/H]=-1 with those for [Fe/H]=-3. The productions of
Be and B are proportional to the abundance of metals, because these species 
result from spallation of fast (resp. ambient) C, N and O interacting with
ambient (resp. fast) H and He. On the other hand, the productions of $^6$Li and 
$^7$Li are almost independent of metallicity, because, for [Fe/H]$<$-1, the Li
isotopes are produced almost exclusively in He+He reactions. Consequences of
these different metallicity dependences are discussed in Sect.~4.3. 

\section{$^6$Li production model}

\subsection{Time evolution of $^6$Li abundance in MPHSs}

We calculated the $^6$Li abundance in the atmosphere of MPHSs by assuming that 
the $^6$Li nuclei produced by interaction of flare-accelerated particles with 
stellar atmospheric matter are instantaneously mixed into the bulk of the 
surface convection zone (SCZ). This is justified because the characteristic 
timescales for variation of the depth of the SCZ are much larger than typical 
convective turnover times (Kim \& Demarque 1996). If the mass contained in the 
SCZ, $M_{cz}(t)$, is a decreasing function of time, which is the case during 
the main sequence evolution (e.g. Richard et al. 2002), the net rate of 
variation of $^6$Li/H at the surface of a MPHS is independent of the variation
of $M_{cz}(t)$, and can be written as
\begin{equation} 
{d \over dt}\bigg({^6{\rm Li} \over {\rm H}}\bigg) = {Q(^6{\rm Li}) 
L_p^\odot(\ge10~{\rm MeV}) f_a(t) \over M_{cz}(t)/(1.4m_p)} - 
\lambda_{D}(t)\bigg({^6{\rm Li} \over {\rm H}}\bigg)~,
\end{equation}
where $Q(^6{\rm Li})$ is the $^6$Li production yield plotted in Fig.~1 
($Q(^6{\rm Li})$$\sim$2~nuclei per erg for $s$=4), 
$L_p^\odot(\ge10~{\rm MeV})$$\sim$10$^{23}$~erg~s$^{-1}$ is the average power 
contained in solar-flare accelerated protons of energy $E$$\ge10~{\rm MeV}$ 
(Ramaty \& Simnett 1991; Ramaty et al. 2000b), $f_a(t)$ is the time-dependent 
luminosity of the flare-accelerated protons which irradiate the MPHS with 
respect to that of the contemporary Sun, $m_p$ is the proton mass, and 
$\lambda_{D}(t)$ is the characteristic rate of $^6$Li loss from the SCZ. This 
last term accounts for $^6$Li stellar depletion by various processes including 
nuclear burning, gravitational settling, rotational mixing, and the possible 
ejection of an unknown fraction of flare-produced nuclei in stellar wind. 
The depletion of H due to the SCZ shrinking is taken as identical to that of 
$^6$Li.

Although young stellar objects are known to present high levels of flaring 
activity in all evolutionary stages from Class I protostars to zero-age main 
sequence (Feigelson \& Montmerle 1999), we show in Sect.~4.2 that $^6$Li 
production on the pre-main sequence can be neglected. This is because (1) the 
duration of the pre-main sequence is much shorter than that of the main 
sequence and (2) the convection zones are deepest during the early stages of
stellar evolution, such that dilution of spallogenic $^6$Li is more important.  
We thus consider here only the production of $^6$Li by stellar flares during 
the main sequence. 

It has been known for many years that the chromospheric and coronal activities 
of dwarf stars are closely related to their rotation (Kraft 1967). This 
relationship results from the generation and amplification of surface magnetic 
fields by a complex dynamo mechanism, whose efficiency depends on the 
interaction between differential rotation and subphotospheric convection 
(Charbonneau \& MacGregor 2001). There exists ample evidence that stellar 
activity depends primarily on two quantities, namely the stellar rotation period, $P_{\rm rot}$, and the effective temperature, $T_{\rm eff}$. The first 
controls how much vorticity is injected into turbulent motions in the 
convective envelope. The second quantity is the best indicator of how deep is 
the SCZ where magnetic fields are generated (Richard et al. 2002). In 
particular, the SCZ characteristics in main-sequence stars of Pop~II are 
extremely close to those in Pop~I stars with the same $T_{\rm eff}$ (e.g. 
Talon \& Charbonnel 2004, Fig.~1). We thus assume that for given 
$T_{\rm eff}$ and $P_{\rm rot}$, both main-sequence star populations 
generate the same surface magnetic fields. 

We also assume that the rate of proton acceleration in stellar flares does not 
depend on metallicity, but only on surface magnetic flux. In solar flares, 
the particle acceleration processes at work in impulsive events are related to 
excitation of plasma waves that are essentially independent of the heavy 
element content of the coronal plasma. Thus, in our model, the scaling of the 
stellar-flare accelerated protons ($f_a(t)$ in eq.~1) only depends on 
$T_{\rm eff}$ and $P_{\rm rot}$, such that it can be estimated using the 
wealth of available data for the stellar activity of Pop~I stars.

Numerous studies have searched for correlations between observable 
manifestations of stellar magnetic activity and rotation. X-ray emission may 
be the most widely available tracer of surface magnetic activity. Although the 
exact mechanism(s) of stellar coronal heating remains poorly understood, it 
leaves no doubt that magnetic fields play the crucial role, such that coronal 
X-ray measurements provide a good proxy for the stellar magnetic flux (Pevtsov 
et al. 2003) and in turn for the flaring activity. We used the extensive study 
of Pizzolato et al. (2003) on the relationship between stellar rotation and 
coronal X-ray emission in Pop~I stars. These authors have selected a sample of 
259 late-type main-sequence stars that were observed with {\it ROSAT}, 
consisting of 110 field stars and 149 stars belonging to young open clusters. 
The stellar rotation periods, which were generally derived from photometric 
measurements, cover the range 0.2$<$$P_{\rm rot}$$<$50 days. Strong 
correlations between the X-ray luminosity, $L_X$, and $P_{\rm rot}$ were found. 
These correlations depend on the $B - V$ color of the stars, which reflects the 
dependence of the surface magnetic fields on the depth of the convection zone. 
To take into account this dependency, stellar magnetic activity is usually 
studied as a function of the so-called Rossby number, 
$R_0$=$P_{\rm rot}$/$\tau_{\rm conv}$, where $\tau_{\rm conv}$ is the convective 
turnover time, which, for main sequence stars, is essentially a function of 
the $B - V$ color only. The 24 stars observed by Asplund et al. (2006) are 
situated in the turnoff region of the Hertzsprung-Russel diagram and cover the 
$B - V$ color range $0.37<B - V<0.49$. Using the database computed with the 
Yale Stellar Evolution Code (Demarque et al. 2004 and references therein), we 
find that earlier in the main sequence life, the stars selected by Asplund et 
al. (2006) were in the color range $\sim$$0.45<B - V<0.60$. Pizzolato et al. 
(2003) obtained the following relationship for the close color range 
$0.50<B - V<0.60$:
\begin{eqnarray} 
L_X & = & 10^{30.1}~{\rm erg~s}^{-1}~~{\rm for}~P_{\rm rot} \le 1.8~{\rm days},
\nonumber \\ 
& = & 10^{30.1} \bigg({P_{\rm rot} \over 1.8~{\rm days}} \bigg)^{-2}
~{\rm erg~s}^{-1}~~{\rm for}~P_{\rm rot} > 1.8~{\rm days}.
\end{eqnarray}
Given the relatively narrow color range of interest, we used for simplicity 
this relation to derive the time-dependent activity level of Pop~I stars, 
instead of a possible more general scaling from the Rossby 
number\footnote{MPHSs probably have much lower X-ray luminosities than those 
given by eq. (2) (see Ottmann et al. 1997), due to the fact that coronal X-ray 
emission is strongly dependent on metallicity, because, for coronal 
temperatures below $\sim$2$\times$10$^7$~K, the bulk of X-ray radiation is 
emitted in lines from heavy elements. For MPHSs, this possibly lower X-ray 
luminosity is not an indication of lower surface magnetic fields.}. 

The time evolution of stellar rotation period during the main sequence
depends on the adopted law for the angular momentum loss of the star. 
Pioneering observations have established that measured rotational velocities 
on the main sequence roughly decrease as the inverse square root of the age  
(Skumanich 1972). This is usually attributed to the loss of angular 
momentum through magnetic stellar winds and corresponds to a loss law of the 
form $dJ/dt \propto \omega^3$, where $\omega$ is the angular velocity 
(Kawaler 1988). Further observations have revealed the existence of more rapid
rotators on the early main sequence than predicted by the above angular 
momentum loss law (see Krishnamurthi et al. 1997 and references therein). 
However, we adopted for simplicity the time dependence
\begin{equation} 
P_{\rm rot}(t)=P_{ZAMS}\sqrt{ {t \over t_{ZAMS}}}~,
\end{equation}
where $P_{ZAMS}$ is the rotation period at the zero-age main sequence (ZAMS),
which we estimated from the recent study of Herbst \& Mundt (2005) (see below).

It is possible, however, that the rotational evolution of Pop~II 
stars has been different from their Pop~I counterparts. The existing data on 
the surface rotation velocity of MPHSs are scarce and not very constraining 
(see Lucatello \& Gratton 2003 and references therein). But it is well known
that radiation-driven mass loss, which carries away angular momentum, is 
metallicity dependent. This is because the main source of radiation opacity is 
provided by metal lines. This effect could lead to higher rotation rate and 
hence higher surface magnetic field and $^6$Li production by flares in MPHSs 
than in Pop~I stars of the same stellar age and initial rotation period 
$P_{ZAMS}$. 

   \begin{figure*}
   \centering
   \includegraphics[width=16.cm]{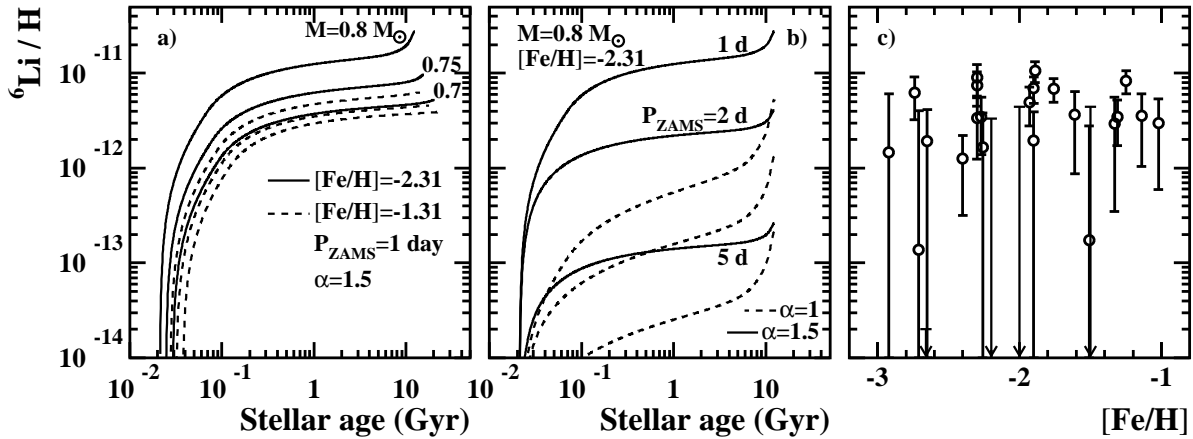}
      \caption{{\it (a)(b)} Calculated evolution of $^6$Li/H on the main 
      sequence of MPHSs. The calculations assume that there is no $^6$Li in the
      stellar atmospheres at the zero-age main sequence and do not take into 
      account $^6$Li depletion. Panel {\it (a)} shows the dependence of 
      predicted $^6$Li/H on the star mass and metallicity, whereas panel {\it 
      (b)} shows the effects of changing $P_{ZAMS}$ (eq.~3) and $\alpha$ 
      (eq.~4). {\it (c)} Observed $^6$Li/H ratios as a function of metallicity. 
      The data are from Asplund et al. (2006), Smith et al. (1998) and Cayrel 
      et al. (1999). The downward-directed arrows correspond to 1$\sigma$
      upper limits, except for the star BD~+09$^\circ$2190 at [Fe/H]=-2.66,
      whose measured isotopic ratio is $^6$Li/$^7$Li=-0.033$\pm$0.027 (Asplund
      et al. 2006).
              }
         \label{Fig2}
   \end{figure*}

Finally to obtain an estimate of $f_a(t)$ (eq.~1), we need to relate the 
coronal X-ray luminosity to the flare-accelerated proton luminosity, 
$L_p(\ge10~{\rm MeV})$. We use the following relation: 
\begin{equation} 
f_a(t) = {L_p(\ge10~{\rm MeV}) \over L_p^\odot(\ge10~{\rm MeV})} 
= \bigg({L_X \over L_X^\odot}\bigg)^\alpha~,
\end{equation}
where the index $\alpha$ accounts for the fact that proton luminosity may
not scale linearly with X-ray luminosity, in the sense that most of the 
energetic protons can be produced by the most powerful flares, whereas heating 
of stellar coronae can essentially be related to less powerful but more 
frequent flares. The relationship between X-ray luminosity and accelerated 
proton fluence has been discussed several times for the young Sun to address 
the role of energetic processes in the early solar system (e.g. Lee et al. 
1998; Feigelson et al. 2002; Gounelle et al. 2006). In these studies, the 
proton luminosity in the young solar nebula is inferred from X-ray observations 
of stellar analogs of the pre-main sequence Sun. The mean characteristic X-ray 
luminosity of these objects is measured to be 
$L_X$$\sim$2$\times10^4L_X^\odot$ and the proton luminosity is estimated to 
be $L_p(\ge10~{\rm MeV})$$\sim$$3\times10^6L_p^\odot(\ge10~{\rm MeV})$. This
corresponds to $\alpha$$\sim$1.5. Feigelson et al. (2002) have 
argued that such a nonlinearity is reasonable because the frequency 
distribution of solar proton events as a function of energy, 
$dN/dE$$\propto$$E^{-1.15}$, is significantly flatter than that of X-ray 
events, $dN/dE$$\propto$$E^{-1.6}-E^{-1.8}$. 

The total luminosity of the Sun's corona in the {\it ROSAT}/PSPC band (0.1--3
keV) ranges from
$\approx$2.7$\times10^{26}$~erg s$^{-1}$ during the quiet phase to 
$\approx$4.7$\times10^{27}$~erg s$^{-1}$ during maximum phase (Peres et al. 
2000). We use the average solar luminosity 
$L_X^\odot$$\simeq$$10^{27}$~erg s$^{-1}$. 

Equations (2--4) allow to estimate the time-dependent luminosity of 
flare-accelerated protons for any Pop~I, main-sequence star in the color range 
$0.5<B - V<0.6$ given its initial rotation period $P_{ZAMS}$. As discussed 
above, this estimate is expected to be valid to first order to Pop~II stars in 
the same color range as well. Clearly, the relation between the 
flare-accelerated proton luminosity and the coronal X-ray luminosity (eq.~4) 
is the most uncertain step of the model, due to our lack of knowledge of the 
heating and particle acceleration processes at work in stellar flares. 

Figures~2a and b show the evolution of $^6$Li/H abundance ratios as the 
function of stellar age calculated from eq.~(1). For these figures, the 
calculations assume no $^6$Li depletion ($\lambda_{D}$=0) and no presence of 
protostellar $^6$Li (for example of cosmic-ray origin) in the stellar 
atmospheres (see below). We took $M_{cz}$ from the evolutionary models of 
MPHSs calculated by Richard et al. (2002, Fig.~1). The age for the beginning of 
the main sequence, $t_{ZAMS}$ in eq.~(3), is also from Richard et al. (2002). 
The dependence of $M_{cz}$ on stellar mass and metallicity can be seen in 
Fig.~2a, where the $^6$Li production is calculated for $P_{ZAMS}$=1~day and 
$\alpha$=1.5. We see that the $^6$Li/H ratio increases with increasing stellar 
mass, because the SCZ becomes shallower. For example, in going from $M_*$=0.7 
to 0.8~$M_\odot$ for [Fe/H]=-2.31, $M_{cz}$ near turnoff is reduced by a factor 
of $\sim$50. The $^6$Li/H ratio also increases with decreasing [Fe/H], because 
the SCZ also becomes shallower as the metallicity is reduced (Richard et al. 
2002; see also Deliyannis \& Malaney 1995).

Figure~2b shows the dependence of $^6$Li production on $\alpha$ and $P_{ZAMS}$. 
The $^6$Li synthesis by flares is less important for $\alpha$=1 than for
$\alpha$=1.5, because the average proton luminosity is lower in the former 
case. The time evolution of the $^6$Li/H ratio is also very different for the 
two values of $\alpha$. For $\alpha$=1, $^6$Li/H is strongly increasing near 
turnoff as $M_{cz}$ becomes shallower. The production of $^6$Li during the 
early main sequence is more important for $\alpha$=1.5, because the 
luminosity of flare-accelerated protons is high enough to compensate the
deeper SCZ. We also see that $^6$Li production strongly depends on $P_{ZAMS}$. 
For $P_{ZAMS}$=1~day, the $^6$Li/H ratio reaches 5.3$\times$10$^{-12}$ and
2.8$\times$10$^{-11}$ at turnoff, for $\alpha$=1 and 1.5, respectively. But for 
$P_{ZAMS}$=5~days, we have only $^6$Li/H$\lsim$3$\times$10$^{-13}$. The latter
value is below the sensitivity of the present $^6$Li measurements, which are
shown in Fig.~2c as a function of metallicity. 

The data shown in this figure are from Asplund et al. (2006), Smith et al. 
(1998) and Cayrel et al. (1999). The data of Asplund et al. (2006) are often 
restricted in the recent literature on this subject (see e.g. Prantzos 2006) 
to the nine detections of $^6$Li at $\geq$2$\sigma$ confidence level. This 
procedure is valuable when the main concern is to discuss the existence of high 
$^6$Li abundances even at low metallicity. But it would have introduced a 
strong bias in our subsequent statistical analysis, especially when analysing 
the dispersion of the data. In the definition of our sample, we thus kept all 
the measured $^6$Li/$^7$Li ratios whatever their values. However,
following Asplund et al. (2006), we excluded from their data the Li-rich star 
HD~106038, which has strong overabundances in Be, Si, Ni, Y and Ba, suggesting 
an enrichment by mass transfer from a companion star. We also did not take into
account the relatively cool ($T_{{\rm eff}}$=5980~K) star HD~19445, whose mass
($M_*$$<$0.7~$M_\odot$) is significantly lower than that of the other stars 
observed by Asplund et al. (2006). As discussed by the authors, this metal-poor
dwarf was only included in the observational program to provide a consistency 
check of the measurements, given that no significant $^6$Li was expected to be
detected. Smith et al. (1998) reported measurements of $^6$Li/$^7$Li isotopic 
ratios with 1$\sigma$ errors in nine single MPHSs. Among them, the four stars 
not observed by Asplund et al. (2006) were included in the data sample: 
HD~74000, HD~84937, BD~+42$^\circ$2667, and BD~+20$^\circ$3603. For HD~84937, 
however, we used the more accurate isotopic ratio measured by Cayrel et al. 
(1999): $^6$Li/$^7$Li=0.052$\pm$0.019. In total, the sample contains 26 stars. 
The $^6$Li/H ratios plotted in Fig.~2c were obtained by multiplying 
$^6$Li/$^7$Li with the measured non-LTE Li abundances.

From a qualitative comparison between Figs.~2b and 2c, it can be 
anticipated that results of the model would not be excluded by the current 
data, provided the distribution of $P_{ZAMS}$ for MPHSs includes both slowly 
($P_{ZAMS}$$\gsim$2~days) and rapidly rotating stars during the early main sequence. To check further the validity of the model, one needs, however, to compare the data to predicted frequency distributions of $^6$Li abundances in MPHSs.
 
\subsection{Frequency distribution of $^6$Li/$^7$Li ratios in MPHSs}

Theoretical frequency distributions of $^6$Li/H values were calculated using 
a Monte-Carlo method. We first generated cubes of $^6$Li abundance values near
turnoff using $P_{ZAMS}$, $M_*$ and [Fe/H] as axes of each cube. We assumed 
the age of the MPHSs to be $\sim$13.5~Gyr (e.g. Richard et al. 2002). For the
stars that are already on the subgiant branch at 13.5~Gyr, we used their 
turnoff age (Richard et al. 2002), which depends on $M_*$ and [Fe/H]. Star
samples with given distributions of rotation rate, mass and metallicity were 
then built by generation of the appropriate random numbers. The $^6$Li surface 
abundance of each generated star was obtained from polynomial interpolation in 
the cubes. The calculated $^6$Li/H values were finally binned to give the
corresponding relative frequencies. 

We estimated the statistical distribution of $P_{ZAMS}$ from the study of 
Herbst \& Mundt (2005). These authors have analyzed available data for the
rotation period of $\sim$500 pre-main sequence and early main sequence stars
belonging to five nearby young clusters: the Orion nebula cluster, NGC 2264,
$\alpha$ Per, IC~2602, and the Pleiades. They found that 50--60\% of young
main sequence stars of solar-like mass (0.4--1.2~$M_\odot$) are rapidly 
rotating ($P_{ZAMS}$$\lsim$2~days). As discussed by Herbst \& Mundt, these 
stars were probably released from any accretion disk locking mechanism very 
early on and thus conserved angular momentum throughout most of their pre-main 
sequence evolution. On the other hand, the remaining 40--50\% stars that are 
more slowly rotating lost substantial amounts of angular momentum during their 
first million years, probably through interactions with their accretion disk. 
We approximated the observed bimodal distribution of stellar rotation period 
by two Gaussian distributions of equal weight, with means and standard 
deviations ($\mu_F$=0.8~day;$\sigma_F$=0.3~day) and 
($\mu_S$=4.3~days;$\sigma_S$=2~days) for the fast and slow rotators, 
respectively. For the fast rotators, a cutoff was introduced at 
$P_{ZAMS}$=0.2~day (Herbst \& Mundt 2005). 

It is likely, however, that the distribution of $P_{ZAMS}$ in metal-poor 
Pop~II stars has been different from that observed in Pop~I stars, due to 
some metallicity effects in the process of star formation, e.g. a possible 
weakening of the magnetic coupling between the forming star and its 
surrounding with decreasing [Fe/H]. Such effects could lead to an enhancement 
of the fraction of rapidly rotating MPHSs in comparison with the proportions 
measured for their Pop I counterparts (see Maeder et al. 1999). 

We assumed the mass distribution of the 26 single stars of the sample to be 
uniform between 0.7 and 0.8~$M_\odot$. This is the approximate mass range 
estimated by Asplund et al. (2006, see Fig.~6) for their star sample, using 
the evolutionary tracks calculated by VandenBerg et al. (2000). 

Figure~3a shows calculated frequency distributions of flare-produced $^6$Li
abundances for 3 stellar metallicities: [Fe/H]=-2.92, -1.02 and -2.02. The 
first two values are the minimum and maximum measured metallicities of the data 
sample. The $^6$Li/H distributions were obtained from the distributions of 
$P_{ZAMS}$ and $M_*$ discussed above, assuming $\alpha$=1.5 (eq.~4) and a 
constant $^6$Li depletion factor for all stars, $D_6$=0.4 dex. This value is 
the approximate average of the main sequence depletion of $^6$Li predicted from 
the so-called T6.09 turbulent diffusion model of Richard et al. (2002, 2005; 
see also Table~6 of Asplund et al. 2006). The bimodal 
character of the $^6$Li/H distributions shown in Fig.~3a 
results from that of the $P_{ZAMS}$ distribution. Thus, rapidly (slowly) 
rotating MPHSs produce $^6$Li/H abundance ratios greater (lower) than 
$\sim$10$^{-12}$. We also see in Fig.~3a that the predicted $^6$Li 
distributions are shifted to higher values as the metallicity decreases. This 
is because the SCZ becomes shallower with decreasing [Fe/H] (see also Fig.2a). 

   \begin{figure*}
   \centering
   \includegraphics[width=16.cm]{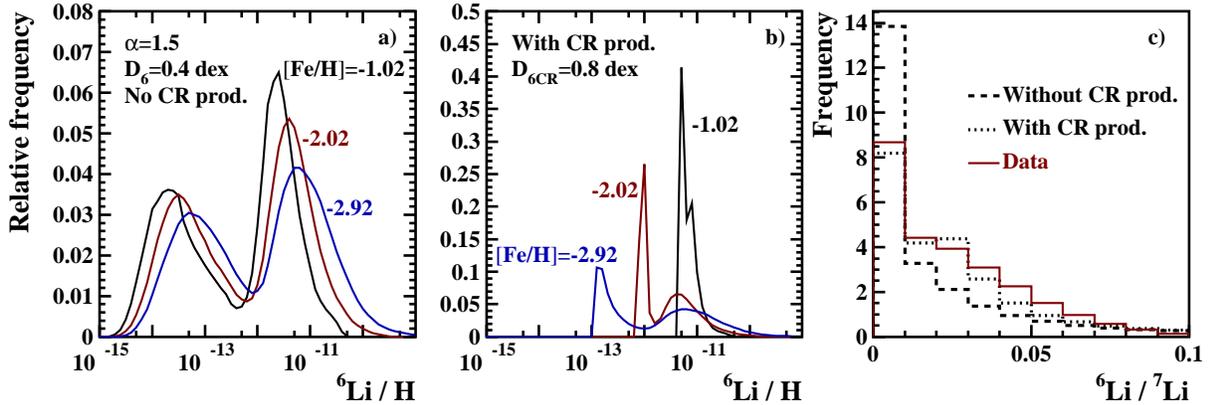}
      \caption{Normalized frequency distributions of {\it (a)} calculated 
      $^6$Li/H ratios for 3 values of [Fe/H] without cosmic-ray $^6$Li 
      production, {\it (b)} the same but with a cosmic-ray $^6$Li contribution, 
      and {\it c}) calculated and observed $^6$Li/$^7$Li ratios averaged over
      metallicity. The calculations assume $\alpha$=1.5 (eq.~4) and depletion 
      factors for flare-produced and cosmic-ray produced $^6$Li of 0.4 and 
      0.8~dex, respectively (see text). 
              }
         \label{Fig3}
   \end{figure*}

We have neglected up to now $^6$Li production by galactic cosmic-ray (GCR)
nucleosynthesis. However, the production of that light isotope by 
interaction of energetic nuclei in GCRs with the interstellar medium is
well established (e.g. Reeves et al. 1970; Vangioni-Flam et al. 2000). In fact, 
galactic chemical evolution models show that $^6$Li production by GCR
interactions can exceed the observed $^6$Li stellar abundances for metallicities
[Fe/H]$\gsim$-2 (Rollinde et al. 2005; Prantzos 2006). This can be accounted 
for by significant $^6$Li depletion at relatively high metallicities. In 
particular, the relatively low abundances measured by Nissen et al. (1999) in 
two galactic disk stars with [Fe/H]$\simeq$-0.6 (HD~68284 and HD~130551; 
$^6$Li/H$\simeq$10$^{-11}$) require a depletion of GCR-produced 
$^6$Li by $\sim$0.8~dex (Prantzos 2006). This depletion factor can result from 
a significant $^6$Li depletion during the pre-main sequence (see Asplund et al. 
2006, Table~6; Piau 2005), in addition to that occuring on the main
sequence. We took into account the GCR-produced $^6$Li from the galactic 
evolution curve shown in Fig.~3 of Prantzos (2006), assuming, for simplicity, 
a metallicity-independent depletion factor $D_{6CR}$=0.8~dex for all 
stars\footnote{The stellar evolution calculations of Richard et al. (2002, 
2005) and Piau (2005) predict an increase of the $^6$Li pre-main sequence
depletion with increasing [Fe/H]. However, the pre-main sequence depletion
factors are considered to be very uncertain (e.g. Proffitt \& Michaud 1989) and 
it is sufficient for the scope of this paper to neglect the possible 
metallicity dependence of $D_{6CR}$.}. $D_{6CR}$ is higher than $D_6$, because
of the additional depletion of GCR-produced $^6$Li on the the pre-main 
sequence. Calculated $^6$Li distributions with the GCR contribution are shown 
in Fig.~3b. The three distributions now appear to be significantly different, 
as a result of the metallicity dependence of the GCR-produced $^6$Li 
component. But we can see from a comparison between Figs.~3a and 3b that for
[Fe/H]$<$-2, the GCR contribution does not significantly modify the
flare-produced $^6$Li/H distributions for the population of fast stellar 
rotators.

In the comparison to the data, both experimental and theoretical $^6$Li/$^7$Li 
ratios were averaged over the metallicity distribution. Although it
necessarily limits the sensitivity of the test of the model, averaging 
over metallicity appears to be unavoidable for the test to be statistically 
significant, given the scarcity of the current data and the relatively large 
experimental uncertainties. 

We show in Fig.~3c a comparison of observed and calculated frequency 
distributions. The data distribution takes into account the experimental 
errors. It was obtained by creating Gaussian distributions whose means were 
the measured isotopic ratios and whose standard deviations were the measured 
1$\sigma$ errors, and then summing them. The resulting distribution was then 
binned into a histogram using a bin width of 0.01, which is comparable to the 
experimental errors. For the data compatible with $^6$Li/H=0, the 
probabilities associated with potential negative values of $^6$Li/H were taken 
into account when calculating the content of the first bin. The theoretical 
distributions were obtained by dividing the $^6$Li abundances calculated for 
each generated MPHS (i.e. as a function of $P_{ZAMS}$, $M_*$ and [Fe/H]), by 
the estimated metallicity-dependent $^7$Li content of that star. We used the 
univariate linear fit given by Asplund et al. (2006):
\begin{equation} 
\log \epsilon_{~^7{\rm Li}}=(2.409 \pm 0.020)+(0.103 \pm 0.010) \cdot 
{\rm [Fe/H]}~,
\end{equation}
where log $\epsilon_{~^7{\rm Li}}$=log($^7$Li/H)+12. 

To test the null hypothesis that the $^6$Li/$^7$Li data sample comes from a 
specific distribution calculated in the framework of the model, we used a 
generalization of the standard chi-square goodness-of-fit test. We chose for 
the test statistic the general $\chi^2$ statistic given by 
Baker \& Cousins (1984) for Poisson-distributed data:
\begin{equation} 
\chi_\lambda^2=2\sum_{i}[(y_i-n_i)+n_i\ln(n_i/y_i)]~,
\end{equation}
where $i$ is the $^6$Li/$^7$Li bin number, $y_i$ is the average number of stars
predicted by the model with a given set of parameters to be in the $i$th bin, 
and $n_i$ is the number of events in the $i$th bin from a given draw of 26 
stars. The probability density function of $\chi_\lambda^2$ was constructed by 
generating random $^6$Li/$^7$Li frequency distributions from a large number of
samples of size $N_{\rm star}$=26, drawn from the population described by the 
model with the given set of parameters. The distributions were limited to the 
seven most significant bins for the test, i.e. $^6$Li/$^7$Li$<$0.07. The 
experimental value, $\chi_{\lambda,{\rm exp}}^2$, was calculated from the data 
distribution using the same values of $y_i$ (i.e. the same model parameters) as 
above. The result of the test was then obtain by deriving from the calculated 
probability density function the probability of observing a value 
$\chi_\lambda^2 > \chi_{\lambda,{\rm exp}}^2$ (see Brandt 1976).

The corresponding probabilities without and with the cosmic-ray $^6$Li 
production were found to be 34.7\% and 99.8\%, respectively. Thus, given the 
usual significance level of 5\%, the null hypothesis "the considered 
$^6$Li/$^7$Li data distribution is correctly described by the model with the 
given parameter values" must not be rejected when cosmic-ray $^6$Li production 
is taken into account. When the contribution of GCR-produced $^6$Li is not 
taken into account, the fraction of slowly rotating stars that produce 
$^6$Li/$^7$Li ratios lower than 0.01 is too high to get a quantitatively good 
agreement with the data (Fig.~3c), even if the overall trend of the observed 
frequency distribution is reproduced. It is noteworthy that a strong 
discrepancy is observed between the data and the model when $^6$Li is supposed 
to be only produced by GCR with a depletion of 0.8~dex. In that case, the 
probability for $^6$Li/$^7$Li to be higher than 0.03 vanishes.

The statistical test described above can also be used to estimate confidence 
intervals for the model parameters. We studied in particular confidence limits 
for $\alpha$ (eq.~4). We found this parameter to be strongly correlated with 
the $^6$Li depletion factor. For the extreme case $D_6$=0 (i.e. no depletion of 
flare-produced $^6$Li) we obtained $\alpha$$>$1.2 with a confidence level of 
68\% (i.e. 1$\sigma$). Thus, a necessary condition for the model to be valid is 
that the stellar-flare-accelerated proton luminosity does not scale linearly 
with the coronal X-ray luminosity.

\section{Discussion}

\subsection{There is no $^6$Li plateau}

We have presented a model of $^6$Li production by stellar flares 
that can explain the recent measurements of $^6$Li in several MPHSs. We have
shown that the measured $^6$Li/$^7$Li frequency distribution can result from a
combination of flare-produced $^6$Li with a protostellar $^6$Li component from
GCR nucleosynthesis. This scenario implies a relatively large dispersion of the
$^6$Li/$^7$Li ratios, which is mainly due to variations in the $^6$Li 
production of each star, rather than variations in the $^6$Li depletion. 

The good agreement observed between data and model predictions suggests a large 
scatter of observational $^6$Li/$^7$Li ratios around mean value. In view of 
this, the existence of a "$^6$Li plateau" from the data of Asplund et al.
(2006), as is almost always reported in the recent literature on this subject, 
appears highly questionable. We note that Asplund et al. themselves carefully 
caution that their $^6$Li data do not necessarily imply the existence of a 
plateau given the observational scatter and the current sensitivity of the 
$^6$Li abundance measurements. This wording is reminiscent of the well-known 
"Spite plateau" for $^7$Li (Spite \& Spite 1982), and thus suggests the same
pre-galactic origin for the two isotopes. 

But we think that it is misleading and
to further demonstrate this point, we performed a statistical test of the
existence of the "$^6$Li plateau". The null hypothesis "there is a plateau" 
(i.e. a moderately tilted averaged linear dependence of $^6$Li/H on metallicity 
with a small data dispersion around the mean value) was tested by performing a 
least squares linear regression to the data of Fig.~2c. From the obtained
minimum $\chi^2$, $\chi^2_{\rm{min}}$=37.4 for 24 degrees of freedom, it can be 
concluded that the hypothesis "there is a plateau" can be rejected at the usual 
significance level of 5\%. This high value of $\chi^2_{\rm{min}}$ reflects a 
significant intrinsic scatter of the data 
($\sigma_{\rm obs}$=3.44$\times$10$^{-12}$ for a 
$^6$Li/H mean value of 3.34$\times$10$^{-12}$), when compared to the measurement
errors. In comparison, the equivalent linear regression performed by Asplund et 
al.(2006) on $^7$Li/H values indicates a very small scatter of 0.033~dex 
around the fit. We checked that similar results were obtained with our somewhat 
enlarged sample, with $\sigma_{\rm obs}$=1.51$\times$10$^{-11}$ for a $^7$Li/H 
mean value of 1.63$\times$10$^{-10}$.

\subsection{$^6$Li production in young stellar objects}

Young stellar objects (YSOs) are sites of very intense flares that are observed 
in X-rays. Recent observations of the Orion Nebula Cluster (ONC) with the 
{\it Chandra} satellite have shown that the X-ray activity of nonaccreting T 
Tauri stars (TTSs) is consistent with that of rapidly rotating main-sequence 
stars, while accreting TTSs are on average less X-ray active (Preibisch et 
al. 2005). These measurements suggest that a similar magnetic dynamo mechanism 
produces the X-ray activity in TTSs and main-sequence stars, although the
correlation between X-ray activity and stellar rotation is not observed in
pre-main sequence stars (Preibisch \& Feigelson 2005; Preibisch et 
al. 2005). Magnetic field generation in YSOs may involve a turbulent convective
dynamo that does not depend on rotation (Barnes 2003).

The flaring activity of YSOs is thought to be responsible for
various high energy processes that are important for the formation of asteroids 
and planets (e.g. Feigelson \& Montmerle 1999). In particular, irradiation of
the inner accretion disk (the so-called "reconnection ring") by
flare-accelerated ions can produce the relatively short-lived radionuclei 
($T_{1/2}$$<$5~Myr) that were present in the solar system when the
calcium-aluminium-rich inclusions were formed (Lee et al. 1998; 
Gounelle et al. 2006 and references therein). In this scenario, the 
flare-accelerated particles are confined by magnetic fields that 
connect the star with the inner part of the accretion disk.

To obtain an upper limit on $^6$Li production in YSOs, we assumed that all 
flare-accelerated particles are impinging on the stellar atmosphere, where they 
produce nuclear reactions in thick target interactions. The level of proton
irradiation was estimated from eq.~(4), with $\alpha$=1.5. We used for the mean
characteristic X-ray luminosity of YSOs the fit as a function of stellar age
obtained by Preibisch \& Feigelson (2005) for ONC stars in the mass range 
0.4--1~$M_\odot$:
\begin{equation} 
\log L_X=30.25 - 0.32 \log 
\bigg({t \over 10^6~{\rm yr}}\bigg)~{\rm erg~s}^{-1}~.
\end{equation}
These X-ray observations were made with {\it Chandra} in the 0.5--8~keV energy 
band. The average solar luminosity in the same energy range is
$\log L_X^\odot$$\sim$26.1~erg~s$^{-1}$ (see Peres et al. 2000). 

The surface $^6$Li abundance should be strongly diluted during the pre-main
sequence, when the convection zones are deepest. We took the time evolution of 
$M_{cz}$ in metal-poor YSOs from Deliyannis \& Malaney (1995, 
Fig.~2). Using their results for [Fe/H]=-2.3 and $M_*$=0.8~$M_\odot$, which 
give the shallowest SCZ for $t$$\gsim$1~Myr, we get from eq.~(1) with 
$\lambda_{D}$=0: $^6$Li/H=4.2$\times$10$^{-13}$ at $t$=10$^7$~yr. The $^6$Li 
abundance will be further depleted before the star reachs the turnoff region. 
Thus, production of spallogenic $^6$Li during the pre-main sequence should not 
significantly contribute to the observed $^6$Li abundances (see Fig.~2c). 

\subsection{$^7$Li, Be and B production by stellar flares}

Deliyannis \& Malaney (1995) have proposed to use the 
$^6$Li/Be and B/Be ratios in the MPHS HD~84937 to discriminate flare-produced 
$^6$Li from $^6$Li of protostellar origin. However, Lemoine et al. (1997) have 
pointed out that the amount of flare-produced Be and B should anyhow be 
negligible with respect to the amount of Be and B observed in this star.

Be and B are produced by spallation from C, N and O, whereas $^6$Li is mainly
produced by the interactions of accelerated $^3$He and $\alpha$-particles with
atmospheric He (Sect.~2). The flare-produced Be/$^6$Li and B/$^6$Li ratios are 
thus proportional to the star metallicity. For the accelerated particle 
spectral index $s$=4, we find the following yield ratios (see Fig.~1):
\begin{eqnarray} 
{Q({\rm Be}) \over Q(^6{\rm Li})} & = & 3.5\times10^{-3} 
{Z \over Z_\odot}~~{\rm and}
\nonumber \\ 
{Q({\rm B}) \over Q(^6{\rm Li})} & = & 0.10 {Z \over Z_\odot}~,
\end{eqnarray}
where $Z/Z_\odot$=10$^{\rm [Fe/H]}$. The observed limit 
$^6$Li/H$<$2$\times$10$^{-11}$ (see Fig.~2c) implies for the flare-produced Be 
and B abundances: 
\begin{eqnarray} 
\bigg({{\rm Be} \over {\rm H}}\bigg)_{\rm Flare} & < & 7\times10^{-14} 
{Z \over Z_\odot}~~{\rm and}
\nonumber \\ 
\bigg({{\rm B} \over {\rm H}}\bigg)_{\rm Flare} & < & 2\times 10^{-12}  
{Z \over Z_\odot}~.
\end{eqnarray}

These limits should be compared with the Be and B abundances measured in MPHSs. 
It is well-known that these abundances are both approximately proportional to 
$Z/Z_\odot$. The following fits to the data have been obtained (see 
Boesgaard et al. 1999 for Be, Duncan et al. 1997 for B):
\begin{eqnarray} 
{{\rm Be} \over {\rm H}} & \simeq & 4\times10^{-11} {Z \over Z_\odot}~~{\rm and}
\nonumber \\ 
{{\rm B} \over {\rm H}} & \simeq & 6 \times 10^{-10} {Z \over Z_\odot}~.
\end{eqnarray}
Thus, the amounts of Be and B produced in flares should always be negligible 
as compared with the observed abundances, as Lemoine et al. (1997) already 
showed for HD~84937. 

This conclusion holds true for $^7$Li as well. For $s$=4, we calculate that
independently of metallicity $Q(^7{\rm Li})$=0.45$Q(^6{\rm Li})$ (Fig.~1). The
measured isotopic ratios $^6$Li/$^7$Li are $<$0.1 (Fig.~3c). Thus, if $^6$Li 
and $^7$Li are similarly depleted during the main sequence (Richard et al. 
2005), flares should contribute less than a few percent to the $^7$Li content 
of MPHSs. 

\subsection{Observational tests of the model}

Flare-produced $^6$Li is most likely to be observed in active MPHSs having a
relatively thin SCZ on the main sequence. The mass $M_{cz}$ of the SCZ 
during the main sequence and at turnoff is a decreasing function of the 
stellar effective temperature $T_{{\rm eff}}$ (e.g. Richard et al. 2002). Thus, 
as discussed by Deliyannis \& Malaney (1995), we expect to observe an increase 
of $^6$Li/$^7$Li as a function of $T_{{\rm eff}}$. Such a relationship, however, 
is also expected from stellar depletion of protostellar Li (Richard et al. 
2005; Deliyannis \& Malaney 1995). Thus, even if a large database of Li 
measurements in MPHSs could allow one to identify a dependence of $^6$Li/$^7$Li 
on $T_{{\rm eff}}$, which is not the case with the current data, this would 
not allow one to unambiguously discriminate between flare-produced and 
protostellar $^6$Li. 

Deliyannis \& Malaney (1995) have proposed more specifically to use 
$^6$Li/$^7$Li as a function of $T_{{\rm eff}}$ on the subgiant branch, in the 
temperature range 6000--6600~K. They argued that if $^6$Li is produced by 
flares, $^6$Li/$^7$Li should decrease with decreasing $T_{{\rm eff}}$ because 
of the increasing $M_{cz}$ past the turnoff, whereas the same ratio should be 
constant if $^6$Li is of protostellar origin, because the preservation region 
of protostellar $^6$Li in subgiant MPHSs is expected to be larger than the 
SCZ. However, Lemoine et al. (1997) have questioned the reliability of this
observational test, which strongly depends on the adopted stellar physics. In 
particular, stellar-evolution models including diffusion and turbulent mixing 
(Richard et al. 2005; Korn et al. 2006) may not predict the same $^6$Li/$^7$Li 
vs. $T_{{\rm eff}}$ relationships.

   \begin{figure}
   \centering
   \includegraphics[width=8.5cm]{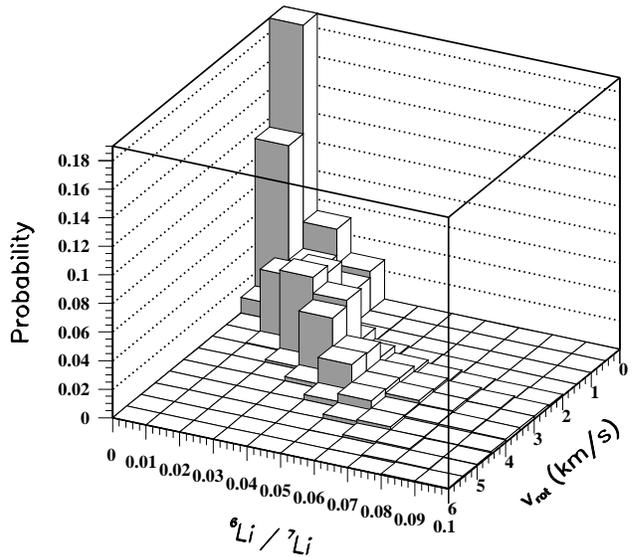}
      \caption{Double differential probability distribution showing the
      correlation expected between $^6$Li/$^7$Li and the stellar rotation
      velocity. The calculations were performed with $\alpha$=1.5, 
      $D_6$=0.4~dex and $D_{6CR}$=0.8~dex.
              }
         \label{Fig4}
   \end{figure}

A more powerful observational test might be provided by the correlation 
expected between the abundance of flare-produced $^6$Li and the stellar 
rotation velocity, which is probably one of the best indicators of the level of
flaring activity that the stars experienced during their main sequence 
evolution. To study this relationship, we introduced the equatorial rotation velocity of each generated star into our Monte-Carlo simulation,
\begin{equation} 
v_{\rm rot}={2 \pi R_* \over P_{\rm rot}}~,
\end{equation}
where the stellar radius $R_*$ at $t$$\sim$13.5~Gyr was obtained from the
Dartmouth stellar evolution code developed by Chaboyer et al. (2001) and 
Guenther et al. (1992)
\footnote{See also URL http://stellar.dartmouth.edu/~evolve/}, and the
rotation period $P_{\rm rot}$ was estimated from eq.~(3). We remind the reader
that eq.~(3) only provides a first-order description of the rotation period
evolution during the main sequence (see Krishnamurthi et al. 1997; Piau 2005). 
Figure~4 shows a calculated distribution of stars as a function of 
$^6$Li/$^7$Li and $v_{\rm rot}$. We used the same parameters as before: 
$\alpha$=1.5, $D_6$=0.4~dex and $D_{6CR}$=0.8~dex (see Fig.~3c). We see that, 
on average, $^6$Li/$^7$Li increases with increasing $v_{\rm rot}$. Thus, the 
calculated mean isotopic ratio is $^6$Li/$^7$Li=0.012 for 
0$<$$v_{\rm rot}$$\leq$2~km~s$^{-1}$ and $^6$Li/$^7$Li=0.045 for 
2$<$$v_{\rm rot}$$\leq$4~km~s$^{-1}$.

We did not take into account in these calculations the possible dependence of 
Li depletion with stellar rotation. Stellar-evolution models with
rotationally-induced mixing predict that $^6$Li depletion increases faster 
than $^7$Li depletion with increasing rotation rate (Pinsonneault et al. 1999).
This effect could lead to a reduction of the $^6$Li/$^7$Li ratio with increasing
$v_{\rm rot}$.

Asplund et al. (2006) have derived the projected rotation velocity 
$v_{\rm rot} \sin i$ of 18 MPHSs from an analysis of several spectral lines 
using 3D LTE model atmospheres. The derived $v_{\rm rot} \sin i$ values range
from 0.4 to 3.8~km~s$^{-1}$. These high estimated $v_{\rm rot} \sin i$ values
would help the flare model for the origin of $^6$Li. However, as discussed by 
Asplund et al., it is possible that part or all of the estimated 
rotational velocities are spoilt by systematic uncertainties due to the adopted 
stellar parameters for the 3D models. We note also that the expected 
correlation should be more difficult to identify with the {\it projected} 
rotation velocity $v_{\rm rot} \sin i$, which of course is lower than the true 
$v_{\rm rot}$ used in Fig.~4.

The relationship between Li abundance and stellar rotation can be more 
easily investigated in young Pop~I stars, which have larger rotational
velocities. Observations of dwarf stars in young open clusters (Soderblom et 
al. 1993; Garc\'ia-L\'opez et al. 1994; Randich et al. 1998) and in the solar 
neighbourhood (Cutispoto et al. 2003) have shown than the largest Li abundances 
are found in rapidly rotating stars with the strongest chromospheric emission. 
These observations are not explained by stellar models of rotation-induced Li 
depletion, which predict on the contrary that fast rotation enhances the mixing 
processes that lead to Li destruction (e.g. Charbonnel et al. 1992; 
Pinsonneault 1997; Piau \& Turck-Chi\`eze 2002). As discussed in Randich et al. 
(1998), it is possible that rapidly rotating, young stars deplete less Li 
because they undergo little angular momentum loss and transport, and hence 
little rotationally-driven mixing, until they reach the ZAMS. The observed 
correlation between $\log \epsilon_{~{\rm Li}}$ and $v_{\rm rot} \sin i$ could 
also be partly due to stellar activity phenomena that may not be properly taken 
into account in Li abundance measurement (Cutispoto 2002; Xiong \& Deng 2005). 
However, Li production by flares could also play a role, although the SCZs of 
these young stars are relatively deep. Our calculations show that the high Li 
abundances found in the fast rotating stars (Li/H$\sim$10$^{-9}$) could be 
produced in flares, only if the spallogenic Li is not too rapidly diluted 
into the bulk of the SCZs. A large fraction of these Li atoms would then be 
$^6$Li.

\section{Summary}

We have developed a model of light element production by stellar flares that 
can explain the recent observations of $^6$Li in several MPHSs near turnoff,
thus avoiding the need for a significant pre-galactic source of this isotope.
We have shown that $^6$Li could be mostly produced by the reaction 
$^4$He($^3$He,$p$)$^6$Li, if, as in solar flares, energetic $^3$He nuclei are 
strongly enriched by a stochastic acceleration process.  

The model is based on our current knowledge of the flaring activity of Pop~I
dwarf stars. Assuming for these high-metallicity objects a power-law 
relationship between the luminosity of flare-accelerated protons and the 
coronal X-ray luminosity, and using a well-established dependence of the X-ray 
luminosity on stellar rotation and in consequence on age, we have constructed 
a simple time-dependent model for the average power contained in stellar-flare
accelerated particles. We then have used this flaring-activity parameter to 
model the production of $^6$Li in the atmospheres of MPHSs.

Theoretical frequency distributions of $^6$Li/$^7$Li in MPHSs near turnoff were 
calculated using a Monte-Carlo method and compared with the data of Smith et 
al. (1998), Cayrel et al. (1999) and Asplund et al. (2006). We took into 
account protostellar $^6$Li produced by GCR nucleosynthesis as an additional 
contribution, which is only significant in stars of [Fe/H]$>$-2. Excellent
agreement was found with the measured $^6$Li/$^7$Li distribution. 

Both stellar depletion factors for flare-produced and cosmic-ray produced 
$^6$Li were treated in a first approximation as metallicity-independent 
parameters. The average values of these parameters were estimated from the
stellar-evolution calculations of Richard et al. (2002, 2005). Prantzos (2006)
has argued that if a pre-galactic source of protostellar $^6$Li is to explain 
the abundances measured in the lowest metallicity stars, the unavoidable 
contribution of GCR-produced $^6$Li at [Fe/H]$>$-2 implies the existence of a 
"fine-tuned and metallicity-dependent depletion mechanism". This "fine-tuning" 
is relaxed in the flare model, because the flare-produced $^6$Li component is 
less dependent on metallicity and less depleted than the protostellar $^6$Li 
contribution from the GCR.

An observational signature of the flare model will be difficult to obtain. We 
propose to seek for a positive correlation between $^6$Li/$^7$Li and stellar 
rotation velocity. We hope that the current development of 3D model 
atmospheres will allow accurate measurements of the projected rotation
velocities of many MPHSs. The predicted increase of $^6$Li/$^7$Li as a function
of $v_{\rm rot}$ is not expected in models of MPHSs that include rotational 
mixing and Li depletion (Pinsonneault et al. 1999).

Further improvements in the model would be necessary to make more detailed
predictions realistic. Such improvements rely on advances in our understanding
of physical processes underlying stellar rotation and flaring activity, as well
as more high-quality measurements of Li isotopic abundances in both Pop~I and 
II stars.

We have shown that the amounts of $^7$Li, Be and B produced in flares should 
always be negligible as compared with the observed abundances of these species 
in MPHSs. $^6$Li may provide a unique tool to study the nuclear processes 
occuring in stellar flares.

\begin{acknowledgements}

We would like to thank Mike Harris for fruitful discussions and his
constructive comments on the manuscript. We are also indebted to Elisabeth
Vangioni, Alain Coc, J\"urgen Kiener, Matthieu Gounelle, and Nikos Prantzos 
for extremely helpful conversations. We finally acknowledge a very valuable 
discussion with Martin Asplund on the $^6$Li data. 

\end{acknowledgements}

\end{document}